\documentclass[conference,A4]{IEEEtran}

 \def\baselinestretch{0.96}

\ifCLASSINFOpdf
  \usepackage[pdftex]{graphicx}
  \DeclareGraphicsExtensions{.pdf,.jpeg,.png}
\else
  \usepackage[dvips]{graphicx}
  \DeclareGraphicsExtensions{.eps}
\fi
\usepackage[cmex10]{amsmath}
\usepackage{array}
\usepackage{fixltx2e}
\usepackage{stfloats}
\usepackage{url}

\let\email\url

\usepackage{times}
\usepackage{alltt}

\def\emquote{\vspace{-.8ex}\begin{quote}\small\em}
\def\endemquote{\end{quote}\vspace{-.8ex}}

\def\zsidebyside{\begin{sidebyside}}
\def\endzsidebyside{\vspace{-2ex}\end{sidebyside}}

\def\eqdef{\mathop{\mathop{=}\limits^{\Delta}}}

\author{\IEEEauthorblockN{Peter T. Breuer}
\IEEEauthorblockA{
Department of Computer Science \\
University of Birmimgham \\
Edgbaston, Birmingham B15 2TT, UK \\
Email: \email{ptb@cs.bham.ac.uk}
}
\and
\IEEEauthorblockN{Jonathan P. Bowen}
\IEEEauthorblockA{
School of Computing, Telecomms and Networking \\
Birmingham City University \\
Millennium Point, Curzon Street \\
Birmingham B4 7XG, UK \\
Email: \email{jonathan.bowen@bcu.ac.uk} \\
URL: \url{http://www.jpbowen.com}
}
}

\title{Empirical Patterns in Google Scholar Citation Counts}

\begin{document}

\maketitle

\IEEEpeerreviewmaketitle

\begin{abstract}
Scholarly impact may be metricized using an author's total number of
citations as a stand-in for real worth, but this measure varies in
applicability between disciplines.
%More recent measures include the h-index and g-index, which
%give more emphasis to the influence of an author's most
%important publications.
The detail of the number of citations per publication is nowadays mapped
in much more detail on the Web, exposing certain empirical patterns.
This paper explores those patterns, using the citation data from Google
Scholar for a number of authors.
\end{abstract}

\section{Background}

%In historic and modern times, the development of knowledge has relied on
%written communications \cite{Van91}.  In academic fields, this has
%typically taken the form of published papers in journals and proceedings
%that are subsequently cited as references in later papers by other
%researchers or by the same author as abbreviations for earlier work.

The speed of transmission and the quantity of knowledge available to
researchers has accelerated dramatically in recent decades with the
advent of the Internet and the World Wide Web.  Whereas, previously,
academic papers were really published only on paper, in journals and
books, now they can be and often are communicated `online'.  That has
led to this body of information on academic activity becoming ever more
comprehensively indexed.

Google, as well as ubiquitously indexing all of the Web, provides
an index of academic publications in particular through its {\em Google
Scholar} website (\url{http://scholar.google.com}) and also provides
access to books though the {\em Google Books} facility
(\url{http://books.google.com}).  It thus has a very complete and
continuously updated collection of academic data available, and is
arguably currently the leading facility of that kind.
Microsoft Academic Search
(\url{http://academic.research.microsoft.com}) provides a
competing database of academic publications online, started at
Microsoft's Beijing research laboratory. While it is not as
complete or up to date as Google Scholar, it does provide better
visualization facilities.

Google Scholar furnishes individual authors with a personalizable page
that presents a list of their own publications and links to the
publications that cite them, with counts of the number of citations per
publication.  The page is generated automatically, but it must be
corrected by hand by the author in order to obtain an accurate record.
Google's automated scanning of online publications works well for
popularly cited works,
because multiple examples in different texts enable Google's automata
to learn to recognize the citation despite differences
in spelling and presentation.

For most productive academic authors, however, there is a `long tail' in
the automatically generated data that consists of those publications with
few or no citations, for which Google's data can be inaccurate and may
well need correction.  Authors with common names may find publications by
other authors with the same or similar names wrongly assigned to their
page, for example.  Google Scholar also confuses publications that have
the superficial appearance of papers (e.g., programme committee
information for conferences) with real papers, and such entries need
to be pruned.  Conversely, publications that are not
represented on the Web at all will not be located by Google, and
must be added in by hand.  Those publications that
have appeared on the Web in slightly different forms ought also to be
merged into a single base entry on the author's page.  An author's
corrections are not cross-checked before appearing online in Google Scholar, 
while corrections made in Microsoft Academic Search are checked before
incorporation, with a delay until the submitted updates appear.

In spite of the more problematic aspects, a Google Scholar page
provides a comprehensive opportunity for administrators to
garner raw statistics on an academic's output, which may affect
prospects for promotion and tenure, and it is therefore likely
that a good proportion of academics are aware of their own
page's existence and have paid some attention to ensuring that
it is reasonably accurate, as well as monitoring it to see how
the numbers grow with time.  The authors of this paper are among
them; we have wondered why the data on our pages looks the way
it does and if there is some underlying pattern to it that we
should be seeing.  This paper points out some of the patterns we
have empirically observed and puts forward a theory as to their
causes.

\section{Citation metrics}

There are a number of metrics in use that have the common aim of
measuring the stature of an academic researcher in their field.
The simplest is an author's total citation count, but it has a
number of drawbacks.  Authors often have a large number of
publications with relatively low numbers of citations that have
had comparatively little influence on their field, so why
account them?  Most researchers of influence have only a small
set of key publications that have been highly cited by their
peers.  For example, Alan Turing \cite{Bow12a} had three key
publications with thousands of citations, each of which have led to
the foundation of important areas of computer science, whereas
some of his works have never been cited (and we will resist the
temptation to skew the statistics by citing them here; the
reader should check Alan Turing's Google Scholar page).
Normally, it is the leading papers by an author that are
accepted as being significant, and the total citations count
obscures the nature of their contribution.  The more highly
regarded metrics weigh the `top end' of of an author's output
more heavily.  The maximal citation count alone is better than
the total for that purpose, but Alan Turing would have his
second and third counts, which are only insignificantly
different from the first, not taken into consideration by that
method.

Google Scholar derives the author's {\em h-index} \cite{Hir05}
and {\em i$_{10}$-index}, which are popular indications of
respectively the depth and breadth of an author's impact,
especially in the field of computer science.  The i$_{10}$-index 
counts the number of the author's works that have received at
least {10} citations.  Looking at the graph of citation counts
per publication arranged in decreasing order left-to-right, it
is the distance out from the left hand side at which the graph
falls below {10} in height.  All other things being equal, the
higher the total number of citations in the graph, the greater
will be the i$_{10}$-index -- that is, assuming that all
researcher's graphs have roughly the same shape.  But are they
the same shape for different researchers?

The h-index is that value of $n$ in the i$_n$ number that
approximately satisfies i$_n = n$.  It is the length of the side
of the biggest square that can be placed under the graph of
citation counts per publication arranged in decreasing order
left-to-right.  So it measures how `fat' the graph is near the
mid-point, in a direction out at 45$^{\rm o}$ to the axes from
the origin.  How might this be expected to change with respect
to total number of citations, for example?
 
%This measures the maximum number of publications by an author
%that have citations greater than or equal to that number.  This
%can be modelled on any given set (which may represent
%publications) using the Z notation \cite{Hen03,Spi01} as
%follows:
%
%\def\hindex{\keyword{h-index}}
%
%\begin{gendef}[X]
%\hindex : \bag X \fun \nat
%\where
%\forall b: \bag X ; n:\nat \spot {} \\
%\t1 \hindex\, b = n \iff {} \\
%\t2 \exists c: \bag X \spot {} \\
%\t3 \# c \geq n \land {} \\
%\t3 c \subseteq b \land {} \\
%\t3 min (\ran c \cup \{0\}) \geq max(\ran (b\setminus c) \cup \{0\})
%\end{gendef}
%
%A bag (or multiset) is a partial function from a set to non-zero
%natural numbers representing the numbers of each of the elements
%in the domain of the function (e.g., the number of citations for
%publications). We are interested in the subset of this that has
%the largest counts of elements. For the h-index measure, the
%size of this set must be at least the minimum count in that set.

These measures need to be treated with caution, since
different fields have significant variation in their patterns of
publication \cite{Sly11}. For example, computer scientists tend
to have a much lower number of co-authors than physicists, whose
co-authors may play a very small role in a highly cited paper. 
Radicchi et al in \cite{RFC} suggest that all citation counts ought to
be normalised with respect to the average citation count per article in 
their field, and they note that such normalized counts seem to follow
the same distribution independent of field.
Nevertheless, the individual numbers for an author's output are
interesting, and we can observe certain empirical relations between
them, which we will elaborate below.

\section{Patterns in citations}

Is there a pattern to the
graph of citation numbers per article for a given author?  On
the default Google Scholar page, the author's articles are
listed with the citation numbers decreasing down the page.
Evidently, there are more entries with, say, 4 citations than
there are with 40, but if one erases the numbers and replaces
them with a plain graph, can one tell how far down the page one
is by the way the graph looks?

The answer is a qualified no -- the graph of citation counts in
decreasing order for a given author in Google Scholar appears to
be `scale invariant': one part of the graph looks just the same
as another part, scaled up or down by a factor.

Evidence of that comes from the result of a test of Benford's
law \cite{Ben} on scale-invariant distributions.  Benford's law says that
the number of individual article citation counts that start with
the digit 1 should be greater than the number that start with 2,
and so on.  The violation of that law for published Iranian
election counts was taken as an indication that the results had
been tampered with \cite{Iran}.  For the two authors of this paper, the
frequencies for the first digits of the citation numbers of
their articles as listed in Google Scholar are shown in
Table~\ref{tab:1}.  They are consistent with the expectation for
scale-invariant distributions -- a decrease in the observed
frequencies from the smaller digits to the larger digits,
ideally down from about 30\% of the total for digit 1 to about
5\% of the total for digit 9.

\begin{table*}[tb]
\caption{Frequencies ($y$-axis) of first digits ($x$-axis) in the
two authors Google Scholar article citations numbers.}
\begin{center}
% \begin{tabular}{@{}cc@{}}
\begin{tabular}{@{}c@{\qquad}c@{}}
\hbox{
\includegraphics[width=0.9\columnwidth]{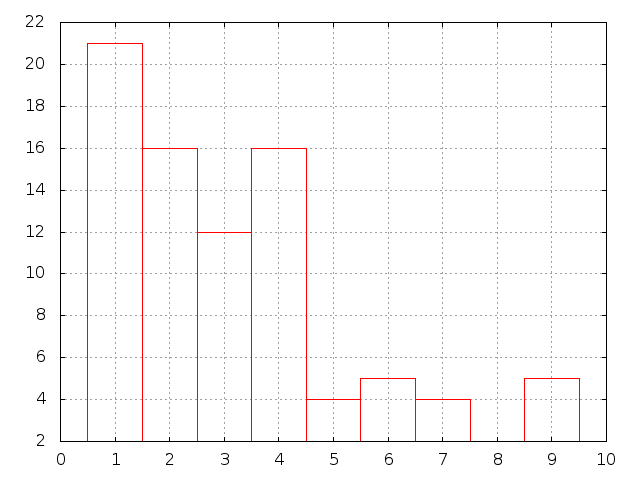}
}
&
\hbox{
\includegraphics[width=0.9\columnwidth]{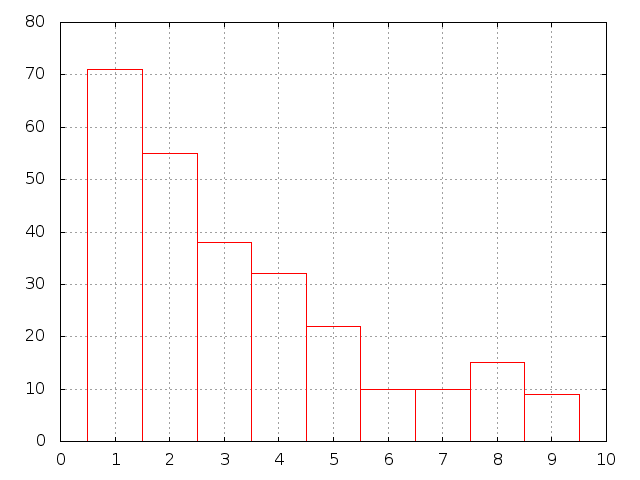}
}
\end{tabular}
\label{tab:1}
\end{center}
\end{table*}

With that evidence for scale-invariance to hand, the question is
what kind of scale-invariant distribution is it. We believe that
the Google Scholar citation counts for articles, laid out in
decreasing order from the article with highest citation count to
the lowest approximately follows an exponential power-law:
\begin{equation}
 c_n \approx c_0\,e^{-P \sqrt{n}}
 \label{eq:c1}
\end{equation}
where $c_n$ is the citation count of the $n$th article in decreasing
order of citation count, and $c_0$
%\begin{equation}
%K=c_0
% \label{eq:K}
%\end{equation}
is the citation count of the most cited article. 
%We will
%use $K$ instead of $c_0$ to emphasise the general natue of some
%formulae, where appropriate.
The factor $c_0$ in \eqref{eq:c1} is chosen so that the
approximation becomes exact at $n=0$.

\subsection{Estimating the multiplier}

Achieving insight into the Google Scholar data for an
author entails getting a good estimate for the multiplier $P$ in
\eqref{eq:c1}.  Below we suggest four increasingly
fit-for-purpose calculations for $P$.  First, putting $n=1$ in
\eqref{eq:c1} gives:
\begin{equation}
P \approx - \ln (c_1/c_0)
\label{eq:P1}
\end{equation}
This is the natural logarithm of the ratio of the citation counts of the
most cited two articles.  This calculation emphasises that $P$ is
related to the initial slope of the graph.  Bigger $P$ means a steeper
initial slope and the `punchier' is the author's best compared to the
rest. Whether that is good or bad shall be left to the reader to decide.

Another estimate comes from putting $n=i_1$ in \eqref{eq:c1},
where $i_1$ is the number of cited articles, the least $n$ such
+that $c_n=0$:
\begin{align*}
i_1 &\eqdef \#\{n\,|\,c_n\ge 1\}\\
    &= 1 + \max\{n\,|\,c_n \ge 1 \}\\
    &= \min\{n\,|\,c_n = 0\}
\end{align*}
and thus $1 \approx c_0 e^{-P\sqrt{i_1}}$, giving rise to
\begin{equation}
P \approx \frac{\ln c_0}{\sqrt{i_1}}
\label{eq:P1a}
\end{equation}
This can be interpreted as the `sharpness' of the roughly
triangular shape formed by the graph of the $\ln c_n$ against
$\sqrt n$. If an author has a highly cited article, this
estimate is larger. But if an author has many hardly-cited
articles, the estimate is lower. An author can trade off a few 
`duds' against an increment in the order of magnitude of the most cited
article. Completely uncited works do not impact this measure at
all.

For the first author of this paper, a good value of $P$ is
empirically about 0.5.  The citations data and the approximating
curve \eqref{eq:c1} for $P=0.5$ are shown together in the left
hand diagram in Table~\ref{tab:2}.  The correspondence is
visually excellent.
 
There is further evidence for the quality of the approximation
\eqref{eq:c1} in the right hand diagram of Table~\ref{tab:2},
which shows the same plot in log-log format, with
$\ln(-\ln\frac{c_n}{c_0})$ being plotted against $\ln n$.  The
approximating curve is transformed by the logarithmic scaling
into the straight line $\ln P + 0.5 \ln n$, and $\ln P$ is where
the line crosses the y-axis, near $-0.7{=}\ln 0.497$.

The estimate \eqref{eq:P1} is $P= 0.4$
(0.401), and \eqref{eq:P1a} gives $P = 0.5$ (0.497).

\begin{table*}[tb]
\caption{Left: Google Scholar citation counts for the first author, 
against the approximation $c_n = c_0 e^{-P\sqrt n}$ with $P=0.5$.
Right: the same data is shown on a log-log graph, with
$\ln(-\ln\frac{c_n}{c_0})$ against $\ln n$.
}
\label{tab:2}
\begin{center}
% \begin{tabular}{@{}c@{~}c@{}}
\begin{tabular}{@{}c@{\qquad}c@{}}
\vbox{\hbox{
\includegraphics[width=0.9\columnwidth]{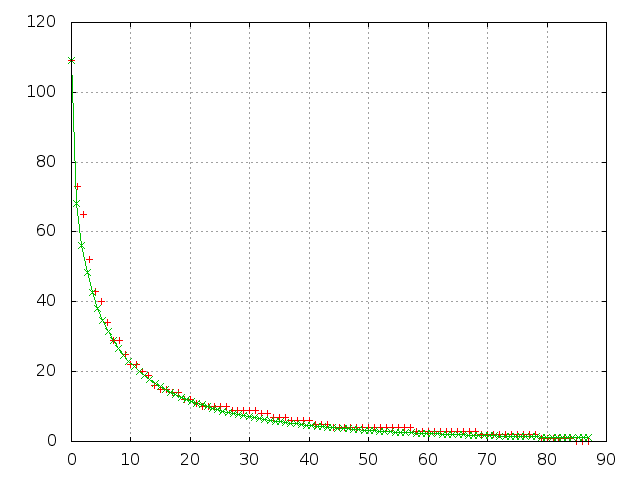}
}}
&
\vbox{\hbox{
\includegraphics[width=0.9\columnwidth]{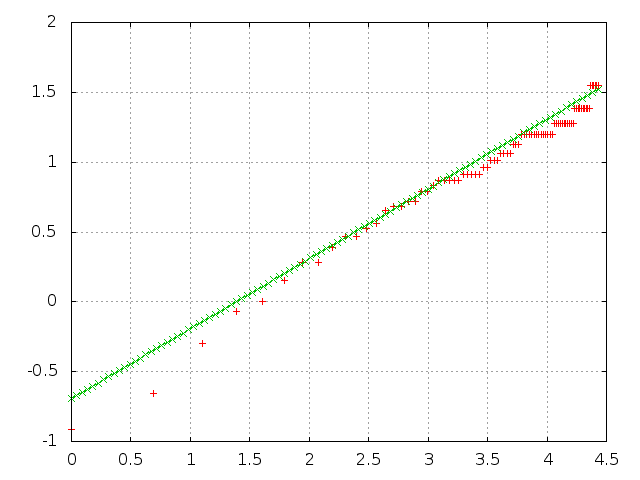}
}}
\end{tabular}
\end{center}
\end{table*}

Finally, here is a holistic estimate for $P$ that takes into account
input from all the data points and which lies between
\eqref{eq:P1} and \eqref{eq:P1a}. It is based on
\begin{equation*}
 \sum_{n=0}^\infty c_n  \approx  2 \frac{c_0}{P^2}
\end{equation*}
where the right hand side is the area under the curve $c_0
e^{-P\sqrt{x}}$ from zero to infinity.  The left hand side is the
total number of citations for
the author, which is also a rough measure of the same area on the plot.
We will call this total $S$ from now on:
\begin{equation*}
S \eqdef \sum_{n=0}^\infty c_n
\end{equation*}
Thus $S \approx  2 c_0/ P^2$, giving the estimate:
\begin{equation}
P \approx \sqrt{2 c_0/S}
\label{eq:P2a}
\end{equation}
For the first author of this paper, \eqref{eq:P2a} yields
$P=0.47$, between the 0.5 posited from the acuteness of the
log/root triangle formed by the citations graph, and the 0.40
from the log of the ratio of the top two citation numbers.  It
is a balanced estimator over the whole set of data, but in
consequence the first few citation numbers (the most cited on
down) are not so well approximated by it.  The effect is visible
in the log-log plot (Table~\ref{tab:2}), where the first few
data points appear off and below the approximating curve
although the tail is well approximated.  On the plot with
unscaled axes, however, the deviation is hardly noticeable.

The fourth estimate of $P$ comes from supposing the value of
$\ln P$ is where a best-fit straight-line approximation on the
log-log graph of citations crosses the y-axis. Formalised in
terms of the covariance and averages of the log-log data, it is
\begin{equation}
\ln P \approx
\text{av} (\ln(\ln \frac{c_0}{c_n})) - \text{av}(\ln
n)~\text{cov}(\ln(\ln\frac{c_0}{c_n}),\ln n)
\end{equation}
where the uncited papers and the very top cited paper are left out
of the reckoning here.  The arithmetic averages of the logarithms are
the logarithms of the geometric means.

% P = geom(ln c0 - ln cn)
%     -------------------
%     (N!)^(1/N) cov

\subsection{Adjusting the shape of the curve for different authors}

We do not yet have an underlying rationale for the term
$\sqrt{n}$ in the empirically observed formula \eqref{eq:c1}.
It seems not to be quite right for some authors, and in general
we would like to suppose that the term is $n^A$ for some
constant $A$ that just happens to be approximately $A=0.5$ in
the case of the first author of this paper. The more general
approximation is:
\begin{equation}
 c_n \approx  c_0 e^{-P n^A}
 \label{eq:c2}
\end{equation}
and $\eqref{eq:c1}$ is  $\eqref{eq:c2}$ with $A=0.5$.  A log-log
plot like that on the right in Table~\ref{tab:2} allows $A$ to
be estimated by the slope of the approximating straight line,
and placing the line by eye on the plot is a good practical
means of positing a value for $A$.

\begin{table*}[tb]
\caption{Left: citation counts for Alan Turing, 
against the approximation $c_n = c_0 e^{-0.95 n^{0.5}}$ (green), and
$c_0 e^{-1.50 n^{0.4}}$ (blue).
Right: the same data on a log-log graph, with
$\ln(-\ln\frac{c_n}{c_0})$ against $\ln n$.
}
\label{tab:3}
\begin{center}
% \begin{tabular}{@{}c@{~}c@{}}
\begin{tabular}{@{}c@{\qquad}c@{}}
\vbox{\hbox{
\includegraphics[width=0.9\columnwidth]{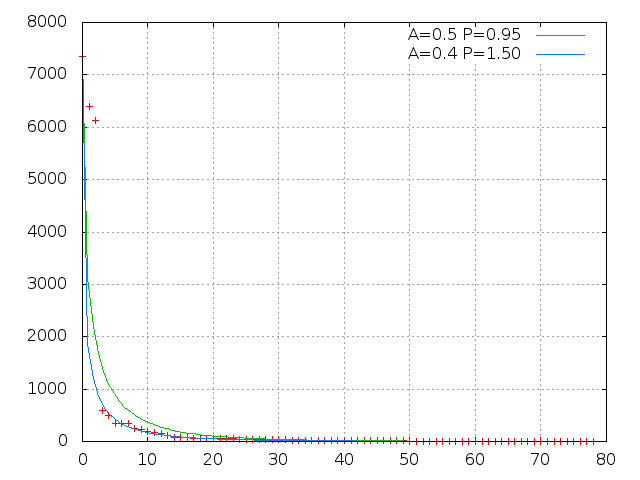}
}}
&
\vbox{\hbox{
\includegraphics[width=0.9\columnwidth]{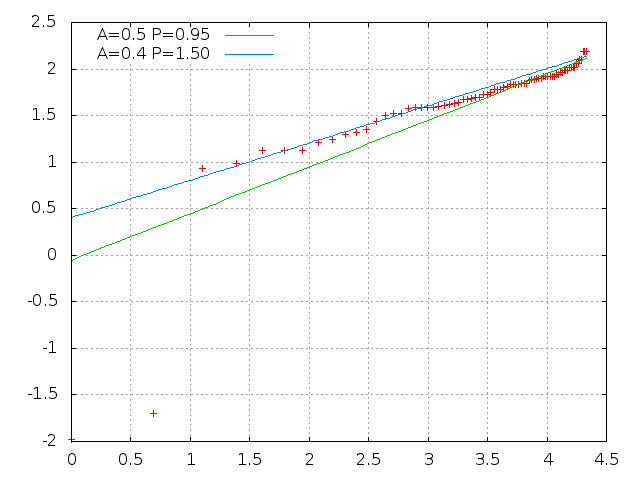}
}}
\end{tabular}
\end{center}
\end{table*}

For Alan Turing's citation data as shown on Google Scholar, we
obtain a better approximation with $A=0.4$ than with $A=0.5$.
The approximation with $A=0.5$ is shown in green in
Table~\ref{tab:3}.
%The best value of $P$ for it is about $P=0.95$.
The approximating curve is high in mid-range, and a little low
farther up-range. That is not too surprising, given the
abnormality of Turing's data. He has three most cited papers of
roughly the same order, and then a fourth and more papers an
order of magnitude less cited (but still enormously highly cited
by most standards).  We could never hope to capture three nearly
equal top papers with the kind of approximation in \eqref{eq:c2}
and $|A|{<}1$ because of the sharp peak that approximation
produces at zero.  Turing's data is more like what one would
expect from three contributors, or three equal careers (indeed,
one of the top three papers is in mathematical biology).  Still
the log-log curve shows clearly that 0.5 is too steep a slope
for the straight line approximation after the first few points.
We do need an approximating term more like
$^{2.5}\kern-5pt\sqrt{n}$ than $\sqrt{n}$ for the tail.

Plotting Alan Turing's data against $A=0.4$ gives the
blue lines in Table~\ref{tab:3}. The log-log graph looks perfect.

How does one estimate $P$ numerically in the general case?

The `area under the curve' argument on the unscaled graph gives
the following approximation when $A$ is the reciprocal of an
integer, $A=\frac{1}{2},\frac{1}{3},\frac{1}{4},\dots$ (but not,
yet, for $A=0.40$):
\begin{equation*}
P \approx  (c_0\,(A^{-1})!/S)^A
%\label{eq:P2}
\end{equation*}
For other values of $A$, we need to replace the factorial
expression using Euler's gamma function:
\begin{equation}
P \approx  (c_0\,\Gamma(1+A^{-1}) /S)^A
\label{eq:P3}
\end{equation}
Varying $A$ then allows approximations to be fine-tuned.

\begin{table*}[tbp]
\caption{Left: citation counts for the second author, 
against the approximation $c_n = c_0 e^{-P n^{0.4}}$ with $P=0.5$.
Right: the same data shown on a log-log graph, with
$\ln(-\ln\frac{c_n}{c_0})$ against $\ln n$.
}
\label{tab:5}
\begin{center}
% \begin{tabular}{@{}c@{~}c@{}}
\begin{tabular}{@{}c@{\qquad}c@{}}
\vbox{\hbox{
\includegraphics[width=0.9\columnwidth]{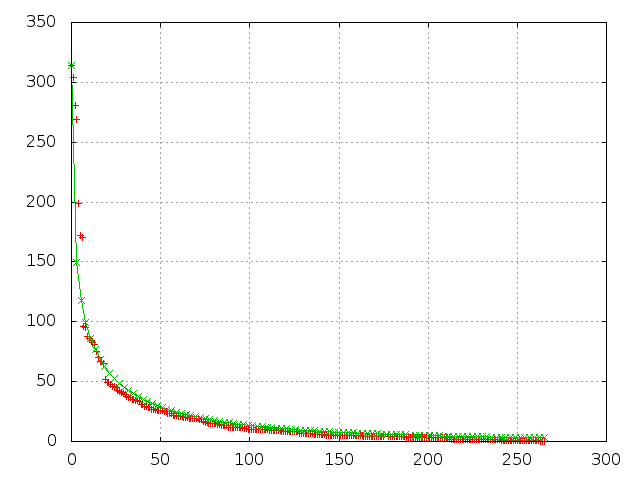}
}}
&
\vbox{\hbox{
\includegraphics[width=0.9\columnwidth]{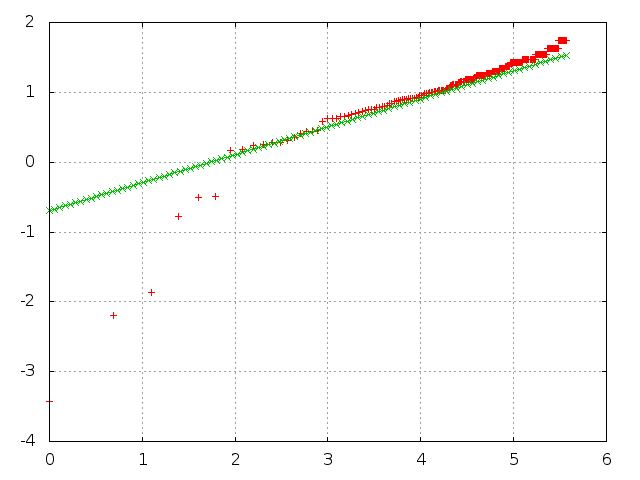}
}}
\end{tabular}
\end{center}
\end{table*}

%1/A = 2.5, so \Gamma(1+1/0.4) = \Gamma(3.5) = 2.5 1.5 \Gamma(1.5)
%                              = 2.5 1.5 0.5 \Gamma(0.5)
%                              = 2.5 1.5 0.5 sqrt(pi)
%                            
%\Gamma(1.4) = 0.8872638175
%\Gamma(1.5) = 0.8862269255

For the second author of this paper, $A=0.4$ also appears to be
better than $A=0.5$. The plots for $A=0.4$ are shown in
Table~\ref{tab:5}, and $P=0.5$ is approximately right for it by
eye.  In this case $\Gamma(1+A^{-1}) = \Gamma(\frac{7}{2}) =
\frac{15}{8} \sqrt{\pi}$, so \eqref{eq:P3} becomes
\begin{equation*}
P \approx \left(\frac{15 c_0}{8S} \sqrt{\pi}\right)^{0.400}
\end{equation*}
which is 0.520 because $c_0/S = 0.0587$ for this author.  The number is
in good agreement with the visually fitted value of $P=0.5$.

%Can we estimate $A$ from first principles?  The classic
%statistical cross-correlation calculation between x- and y-
%ordinates in the log-log graph formally gives the slope of the
%straight line minimising the least square errors.  But that
%gives weight to the data points at the left of the plots, where
%there may be anomalous values that do not fit the general
%pattern, as is the case with Turing's data.  One should
%preferentially fit the data in the tail to obtain $A$.  Large
%vertical errors in the (few) points to the left in the graph
%are hardly noticeable, because the slope of the data and the
%approximating curve is so large there.  Badly approximated data
%points there look displaced horizontally by a small amount, not
%displaced vertically by a large amount.

\subsection{Classical citation-based measures of worth}

The $i_{10}$, $i_{20}$ measures promoted on Google Scholar and
elsewhere are defined by $c_{i_k-1}\ge k >c_{i_k}$, so:
\begin{align*}
10 &\approx c_{i_{10}}\approx c_0 e^{-P (i_{10})^A}\\
20 &\approx c_{i_{20}}\approx c_0 e^{-P (i_{20})^A}
\end{align*}
Thus
\begin{align*}
\ln(10/c_0) &\approx -P (i_{10})^A\\
\ln(20/c_0) &\approx -P (i_{20})^A
\end{align*}
and
\[
\ln(10/c_0)/\ln(20/c_0) \approx (i_{10}/i_{20})^A
\]
or
%\[
%i_{10}/i_{20} \approx (1 + \ln 2/\ln(10/c_0))^{-1/A}
%\]
\begin{equation}
i_{10}/i_{20} \approx {}^A\kern-2pt\sqrt{\ln (c_0/10) / \ln (c_0/20)}
\label{eq:i1020}
\end{equation}
For the first author of this paper and $A=0.5$, the predicted ratio
\eqref{eq:i1020} is 1.98, against the real ratio
$i_{10}/i_{20} =  2.08$.  The estimate is very close.
For the second author of this paper and $A=0.4$, the predicted ratio
\eqref{eq:i1020} is 1.75 and $i_{10}/i_{20} = 1.69$ in reality, again
very close.
For Alan Turing, with $A=0.4$  the predicted ratio is
\eqref{eq:i1020} is 1.32 and $i_{10}/i_{20} =
1.44$ in reality, somewhat less close, but Turing's numbers are
extraordinary. It is striking, however, that the prediction adjusts
to approximately match the author in every case, despite their 
differences.

% for h, i10/h = (ln(c0/10)/ln(c0/h))^(1/A)
%        (i10/h)^A = ln(c0/10)/ln(c0/h)
%        (i10/h)^A ln(c0/h) = ln(c0/10)
%        f(h) = ln(c0/10)
% where f(h)  = ln(c0/h) (i10/h)^A

We can derive a relationship between $i_{10}$ and the h-index,
using $i_h \approx h$ and replacing both $i_{20}$ and $20$ by $h$
in \eqref{eq:i1020}. Then
\begin{equation}
 %\ln(c_0/h) (i_{10}/h)^A \approx \ln(c_0/10)
 (c_0/h)^A \ln(c_0/h)  \approx (c_0/i_{10})^A \ln(c_0/10) 
 \label{eq:const}
\end{equation}
For the first author and $A=0.5$, the function on the left is
$(109/h)^{1/2}\ln(109/h)$, and the constant on the right is 4.8,
with solution $h\approx 16.67$.  The truth is that the author's h-index is
15 -- but it is only one citation away from 16.

The number on the right in \eqref{eq:const} is fairly constant across a
range of i-indices for a given author. For the first author of this
paper, it is in the range 4 to 5 up to about $i_{50}$. For the second
author of this paper, it is in the range 5 to 6 up to about $i_{100}$.
A larger number means that the ratio $c_0/h$ is larger, and the h-index is a
smaller fraction of the peak (`most-cited') number.

\section{Underlying causes}

%\begin{table}[h]
%\caption{Article citation count bar chart for the second author of this
%paper.}
%\label{tab:bar}
%\begin{center}
%\vbox{\hbox{
%\includegraphics[width=0.9\columnwidth]{bowenout.png}
%}}
%\end{center}
%\end{table}

How can we explain these observations?

Plotting the number of articles for which the number of citations
$C$ falls in $x \le \ln C / \ln c_0 < x + dx$ against
$x = \ln C / \ln c_0$ for $C\ge 1$ gives a curve 
with mean $\mu$ and standard deviation $\sigma$. That is,
$(\ln C/\ln c_0 - \mu)/\sigma$ looks like a 
random variable with mean 0 and standard deviation 1. 
%The plot for the second author of this paper is shown in Table~\ref{tab:bar}.
Moreover, in the data sets we have looked at, the mean $\mu$
and standard deviation
$\sigma$ are
nearly the same (approximately 0.2) in every case. For the first and
second authors, and Alan Turing, the mean and standard deviation
pairs $(\mu,\sigma)$
are respectively (0.251331, 0.216663), (0.217722, 0.211669),
(0.190153, 0.196518).
Say:
\begin{equation*}
\sigma = \mu = \lambda
%\label{eq:lambda}
\end{equation*}
We may take $\ln C$ to be normally distributed with mean $\mu = \lambda\ln
c_0$ and standard deviation $\sigma = \lambda \ln c_0$.
By the standard statistics of log-normal
distributions, one expects citations $C$ to have mean $m$ and standard
deviation $s$ where
\begin{align*}
m&= c_0^{2\lambda}&
s&= c_0^{2\lambda}\sqrt{e^{(\lambda\ln c_0)^2} - 1}
\end{align*}
which allows the parameter $\lambda$ to be conveniently
estimated from $s/m$.  For the first and second authors, and
Alan Turing, $\lambda$ estimated this way is 0.264, 0.250, 0.209
respectively.

We have generated sets of fake citation data using a normally
distributed random variable for $\ln C/\ln c_0$ with both
mean and standard deviation equal to $\lambda$. That is, $C =
e^{(\lambda +\lambda X)\ln c_0}$ where $X$ is a normally
distributed random variable with mean 0 and standard deviation
1.  The generated data looks like a real citations count list,
and ordering it in descending order $c_0$, $c_1$, \dots and
plotting it in log-log as in Table~\ref{tab:2}-\ref{tab:5} (with
$\ln(-\ln(c_n/c_0))$ against $\ln n$) shows that a straight line
approximation is appropriate.  The slope of the line is the $A$
of the approximation \eqref{eq:c2}, and the slopes manifestly
cluster around $A=0.4$, but can vary between 0.2 and 0.6. 

Placing the best fit line via least squares minimisation of the
errors on the log-log plots (excluding the first citation
number) gives the slope as the covariance between x- and
y-ordinates of the data points. That gives the following
estimates of slope $A$ for $N\,{=}\,200$ datapoints.  The average is taken over
100 generated datasets for each value of $\lambda$ listed:

\begin{center}
\begin{tabular}{|l|ll|}
\hline\vspace{.3ex}
$\lambda$ & slope $A$ & \begin{tabular}{@{}l@{}}standard \\
deviation\end{tabular}\\ 
\hline
0.2\hbox to 0pt{\vbox to 2ex{}} &0.396119&0.075180\\
0.25&0.405471&0.077208\\
0.3 &0.411451&0.080782\\
0.35&0.384616&0.078864\\
0.4 &0.400892&0.087734\\
0.45&0.413907&0.090714\\
\hline
\end{tabular}
\end{center}
This is empirical support for the approximations with power exponents
$A{=}0.4$ (exceptionally 0.5, in the case of Table~\ref{tab:2})
as in Tables~\ref{tab:2}-\ref{tab:5}. The value of $\lambda$ has
insignificant effect.

This value of $A$ arises naturally. It is the slope of the best-fit
line to the logarithm of normally distributed data with equal mean and
standard deviation that
has been ranked in decreasing order against the logarithm of the ranking
position.\footnote{
The slope is that of $\ln(x_0 - x_n)$ against $\ln n$ for a standard
normal variable $X$, with the $N$ observations arranged in decreasing order
$x_0,\,x_1,\,\dots$. Theory says that the $x_n$ are positioned about
where the quantiles function $q(\rho) = x \Leftrightarrow prob(X{<}x) =
\rho$ says they should be for $\rho = (n+0.5)/N$, at
$x_n \approx q((n+0.5)/N)$.  In particular, the
cumulative density function for the maximum of the $N$ observations,
$\text{prob}(x_0{<}x)$, is $(\text{prob}(X{<}x))^N$, the $N$th
power of the cumulative density function of an individual observation,
and the position of the maximum observation is expected to be $x_0
\approx x : \text{prob}(X{>}x) = 1/(2N)$; i.e., $x_0 \approx q(1-1/(2N))$.
For the normally distributed
standard variable $X$, that is $x_0 \approx \frac{1}{2} \sqrt{2 \ln
N}$ asymptotically, applying classical mathematical analysis
to the integral that defines $q$.
The minimum is expected to be the same distance in the other
direction. Thus the slope $\ln(x_0 - x_{N{-}1})/\ln N$ is approximated by
$\ln \sqrt{2 \ln N}/ \ln N$, or $\ln(\ln N^2)/\ln N^2$.
}
But the precise value depends on the number of datapoints 
$N$ as follows:

\begin{center}
\begin{tabular}{|l|ll|}
\hline\vspace{.3ex}
$N$ & slope $A$ & \begin{tabular}{@{}l@{}}standard\\
deviation\end{tabular}\\ 
\hline
100\hbox to 0pt{\vbox to 2ex{}}     & 0.463705 &0.105916\\
1000    & 0.363892 &0.052899\\
10000   & 0.279002 &0.032274\\
100000  & 0.233939 &0.017043\\
1000000 & 0.204701 &0.012906\\
\hline
\end{tabular}
\end{center}
The slope $A$ slowly decreases to zero
with increasing $N$.  The measurements in the table above vary from
0.95 to 0.85 of a predicted bounding asymptote $\ln(\ln N^2)/\ln
N^2$.

The approximation \eqref{eq:c2} is 
compatible with observations from a log-normal distribution. The
rate $A$ in the exponent varies
slightly according to the number of an author's publications, but is otherwise
stable across authors.  The number $P$ determines how exceptional are
the most cited articles with respect to the body of work of an
author.  For the authors of this paper, $P=0.5$ is about right.  For
Alan Turing $P=1.5$ is indicated, highlighting the extra 
significance of his three top papers relative to the rest of his (also highly
significant) work.

The numbers $1 - i_n/N$ provide a direct measurement of the 
cumulative density function $\text{prob}(C<n)$ for the random variable $C$ underlying the
citations counts.  The derivative is the probability density function.
The logarithm of citation count ($\ln C$; the x-axis of the density
function) looks distributed like a Poisson distribution (the y-axis).  A
normal distribution with equal mean and standard deviation is a fair
approximation to Poisson, and is what we have used in our analysis.

A Poisson distribution represents low probability events (citations!)
accruing in several equal sized slots over time. After a while, most slots
have the average number of events in, while a very few have none, and a
very few have a large number of events in.
The situation here is that instead of seeing, say, $k$ slots with $c$
events as expected according to a Poisson distribution, we are
seeing $k$ articles with $\ln c$ citations in. We do not have good
insight into that from the publications and citations point of view.
Perhaps it means that the average `intrinsic worth' of an article is
distributed by a Poisson process, but that articles accumulate
citations according to the exponential of their worth. Citation begets
citations, in other words.

Radicci et al report in \cite{RFC} that citation counts relative to the
average count in a field follow (the same) log-normal distribution
irresepctive of field.  We also see log-normal distribution, but
within a single author's output, so perhaps the results of \cite{RFC}
apply when one considers an author as defining their own academic field
of study.  We normalize with respect to the maximal citation count, and
see equal mean and standard deviation (in logarithm), while Radicci et
al normalize with respect to the mean citation count and see mean equal
and opposite in sign to twice the variance (in logarithm).  We believe
these relations are reflections of the same underlying reality. While
Radicci et al sought to quantify an article's worth irrespective of the
field it is published in, we have reduced the question of
an individual author's impact to three parameters, $c_0$, $A$
and $P$, which predict the curve of citation counts.  How these
parameters are distributed across and within academic fields remains to
be discovered.

% what would produce quantile function q(n/N) = c0 e^-Pn^A
% q(p)       = c0 e^{-PN^Ap^A} 
% q(0.5/N)   = c0 e^{-P0.5^A}     large
% q(1-0.5/N) = c0 e^{-P(N-0.5)^A} small
% q(p) = x : prob(X > x) = p = I(x,oo) f(x) dx
% dp/dx = - f(x)
% q'(p) dp/dx = 1
% 1/q'(p)    = - f(x)
% q'(p)      = c0 (-PAN^Ap^(A-1)) e^{-Pn^Ap^A} = (-PAN^Ap^(A-1)) q(p)
% f(x)       = 1/(c0 P A N^A p^(A-1) e^{-Pn^Ap^A})
%            = 1/( P A N^A p^(A-1) q(p))
%            = 1/( P A N^A p^(A-1) x)
% ln(x/c0)   = -PN^A p^A
% -(ln(x/c0)/PN^A) = p^A
% ln(-ln(x/c0)/PN^A) = A ln p
% (1/A) ln(-ln(x/c0)/PN^A) = ln p
% p = (-ln(x/c0)/PN^A)^(1/A)
% so
% f(x)       = 1/(P A N^A p^(A-1) x)
%            = 1/(P A N^A (ln(c0/x)/PN^A)^(1-1/A) x)
%            = K (-ln(x/c0))^(1/A-1) / x
%f(c0-x)     = K (-ln(1-x/c0))^(1/A-1) / (c0 - x)
%f(c0-x)     = K ((x/c0) + (x/c0)^2/2 + ...)^(1/A-1) / c0(1 - x/c0)
%            = K (x/c0) (1 + (x/c0)/2 + ...)^(1/A-1) / c0(1 - x/c0)
%            = K (x/c0) (1 + (1/A-1)(x/c0)/2) / c0(1 - x/c0)
%            = K (x/c0) (1 + (1/A-1)(x/c0)/2) (1 + x/c0 + ...) / c0
%            = K (x/c0) (1 + ((1/A-1)/2 +1)(x/c0)) / c0
%            = K (x/c0) (1 + ((1/A+1)/2)(x/c0)) / c0
%            = K (x/c0) (1 + ((1/A+1)/2)(x/c0)) / c0

\section{Conclusion}
\label{conclusion}

This paper has noted a mathematical pattern with respect to
citation counts for publications of academic authors. When the
citation counts per article are laid out in decreasing order,
they follow the law
\[
  c_n \approx c_0 e^{-P n^A}
\]
for an appropriate multiplier $P$ and rate $A$ fitted to an
individual author.  In practice $A$ ranges from 0.4 to 0.5, and
is lower for higher publication count $N$, decreasing as
$\ln(\ln N^2)/\ln N^2$.  The pattern is compatible with
observations of a log-normal random variable, the exponential
of a normal random variable with equal mean and standard
deviation.
%In particular, the most cited article is distributed
%as the extreme outlier from such a distribution, and its
%logarithm is expected to grow with $\sqrt{\ln N^2}$.  For both
%authors of this paper, the ratio of real to expected value here
%is about 1.5, but for Alan Turing the ratio is about 2.8.

Recognizing these empirical patterns and modelling them allows 
more meaningful metrics to be developed.

Further patterns could be explored among, for example,
temporal information based on the year of publication and
citation of papers \cite{Jon13}. More visualization would also
be possible \cite{Bow13a,Bow13b,Bow12b,Nel06}. Communities of authors
\cite{Bow11}, including their development and demise, could also be
investigated for patterning anomalies.

\paragraph*{Acknowledgements}
Jonathan Bowen is grateful for financial support from Museophile
Limited.  Google Scholar was used to provide publication data
for this paper. 
%The Z notation in this paper has been type-checked using the
%\fuzz\ type-checker \cite{Spi08}.

Peter Breuer is grateful to Richard Gill of the University of
Leiden for conversations that laid bare the statistical
analysis.

\bibliographystyle{plain}
\bibliography{cp14}

\end{document}